\documentclass[12pt]{article}

\usepackage[margin=1in]{geometry}
\usepackage{CJKutf8, hyperref, orcidlink, amsmath, amssymb, amsthm}
\usepackage{pgfplots}
\pgfplotsset{compat=newest}

\DeclareMathOperator{\tr}{tr}

\begin{document}

\title{Entropy of small subsystems in thermalizing systems}

\author{Yichen Huang (黄溢辰)\orcidlink{0000-0002-8496-9251}\\
Santa Clarita, California 91350, USA\\
\href{mailto:huangtbcmh@gmail.com}{huangtbcmh@gmail.com}}

\begin{CJK}{UTF8}{gbsn}

\maketitle

\end{CJK}

\begin{abstract}

We study the entropy of small subsystems in thermalizing quantum many-body systems governed by local Hamiltonians. Assuming the eigenstate thermalization hypothesis, we derive an analytical formula for the von Neumann entropy of equilibrated subsystems. This formula reveals how subsystem entropy depends on the microscopic parameters of the Hamiltonian and the macroscopic properties of the initial state. Furthermore, our results provide a theoretical explanation for recent numerical findings by Maceira and L\"auchli, obtained via exact diagonalization.

\end{abstract}

\section{Introduction}

In isolated quantum many-body systems, the entropy of small subsystems offers valuable insights into how local degrees of freedom reach equilibrium. While the eigenstate thermalization hypothesis (ETH) \cite{Deu91, Sre94, RDO08, DKPR16, Deu18} provides a framework for understanding thermalization, the precise relationship between subsystem entropy and the microscopic details of the Hamiltonian remains an open question.

Assuming the ETH, we derive a closed-form analytical expression for the von Neumann entropy of small subsystems at equilibrium. Specifically, we demonstrate how the equilibrium entropy depends on the microscopic parameters of the Hamiltonian and the energy distribution of the initial state. Our results provide a theoretical explanation for recent numerical findings by Maceira and L\"auchli \cite{ML24}, which uncovered intriguing finite-size effects in subsystem entropy using exact diagonalization.

This study presents an analytical approach for interpreting and predicting the entropy behavior of small subsystems in thermalizing systems, deepening the theoretical understanding of subsystem properties at equilibrium.

\section{Theory}

Consider a constant-dimensional hypercubic lattice of $N$ sites, where each lattice site has a spin. The dimension of the Hilbert space is $d=d_\textnormal{loc}^N$, where $d_\textnormal{loc}$ (a constant) is the local dimension of each spin. The system is governed by a (not necessarily translation-invariant) local Hamiltonian
\begin{equation}
H=\sum_iH_i.
\end{equation}
The sum is over lattice sites. Each term $H_i$ acts non-trivially only on a set of spins contained in a constant-radius neighborhood of site $i$. Assume without loss of generality that $\tr H_i=0$ (traceless) for all $i$ so that the mean energy of $H$ is $\tr H/d=0$. Suppose $H$ is extensive in that
\begin{itemize}
\item $\|H_i\|\le1$ for all $i$.
\item For any site $i$, there exists a site $j$ in a constant-radius neighborhood of site $i$ such that $\|H_j\|\ge c$ for some constant $c>0$.
\end{itemize}

Suppose that the initial state $|\psi(0)\rangle$ has exponential decay of correlations. This includes all product states (each spin is disentangled from all other spins), whose correlation length is zero. Let $A$ be a subsystem of $O(1)$ spins and $\bar A$ be its complement (rest of the system). Let
\begin{equation}
|\psi(t)\rangle=e^{-iHt}|\psi(0)\rangle,\quad\psi(t)_A:=\tr_{\bar A}|\psi(t)\rangle\langle\psi(t)|
\end{equation}
be the state and its reduced density matrix of $A$ at time $t\in\mathbb R$. A thermalizing system equilibrates in that at long times, $\psi(t)_A$ becomes almost independent of time, i.e., its temporal fluctuation vanishes in the thermodynamic limit. The equilibrated $\psi(t)_A$ is given by
\begin{equation}
\psi_A^\infty:=\lim_{\tau\to\infty}\frac1\tau\int_0^\tau\psi(t)_A\,\mathrm dt.
\end{equation}

Suppose that $\langle\psi(0)|H|\psi(0)\rangle=0$ so that the system has the same energy as the infinite-temperature state. Let $I_A$ be the identity operator on and $d_A$ be the dimension of the Hilbert space of subsystem $A$. Assuming the ETH, Eq.~(15) of Ref.~\cite{main} implies that
\begin{equation} 
\|I_A/d_A+cX_A-\psi_A^\infty\|_1=O(1/N^2),
\end{equation}
where $\|\cdot\|_1$ denotes the trace norm, and
\begin{gather}
c:=\frac{\tilde v-v}{2N\tilde v^2},\quad v:=\frac{\langle\psi(0)|H^2|\psi(0)\rangle}N,\quad\tilde v:=\frac{\tr(H^2)}{Nd},\label{eq:c}\\
X_A:=\frac1d\left(\frac{\tr(H^3)\tr_{\bar A}H}{\tr(H^2)}+\frac{\tr(H^2)I_A}{d_A}-\tr_{\bar A}(H^2)\right).\label{eq:X}
\end{gather}
Let $\lambda_1,\lambda_2,\ldots,\lambda_{d_A}$ be the eigenvalues of $X_A$. The eigenvalues of $\psi_A^\infty$ are $1/d_A+\mu_j$, where $\mu_j=c\lambda_j+O(1/N^2)$ for $j=1,2,\ldots,d_A$. Since $|\mu_j|=O(1/N)$ for all $j$, the von Neumann entropy of $\psi_A^\infty$ is 
\begin{multline} \label{eq:inter}
S(\psi_A^\infty):=-\tr(\psi_A^\infty\ln\psi_A^\infty)=-\sum_{j=1}^{d_A}(1/d_A+\mu_j)\ln(1/d_A+\mu_j)=\ln d_A-\frac{d_A}2\sum_{j=1}^{d_A}\mu_j^2+O(1/N^3)\\
=\ln d_A-\frac{c^2d_A}2\sum_{j=1}^{d_A}\lambda_j^2+O(1/N^3)=\ln d_A-c^2d_A\tr(X_A^2)/2+O(1/N^3),
\end{multline}
where we used $\sum_j\mu_j=0$. Plugging Eq.~(\ref{eq:X}) into this equation, we obtain our main result:
\begin{equation} \label{eq:main}
S(\psi_A^\infty)=\ln d_A-\frac{c^2d_A}{2d^2}\tr\left(\left(\frac{\tr(H^3)\tr_{\bar A}H}{\tr(H^2)}+\frac{\tr(H^2)I_A}{d_A}-\tr_{\bar A}(H^2)\right)^2\right)+O(1/N^3).
\end{equation}

\section{Example}

In this section, we apply Eq.~(\ref{eq:main}) to a particular example and compare it with recent numerical results.

Consider a chain of $N$ qubits (spin-$1/2$'s) governed by the Hamiltonian
\begin{equation}
H=\sum_{i=1}^N\sigma_i^z\sigma_{i+1}^z+h_x\sigma_i^x+h_z\sigma_i^z
\end{equation}
with periodic boundary conditions, where $\sigma_i^x,\sigma_i^z$ are the Pauli $x$ and $z$ matrices at site $i$. Suppose that subsystem $A$ consists of the first three qubits so that $d_A=8$. A straightforward calculation shows that
\begin{align}
\frac{\tr(H^2)I_A}{dd_A}-\frac{\tr_{\bar A}(H^2)}d&=-\frac2{d_A}\big(h_x^2\sigma_1^x\sigma_2^x+h_x^2\sigma_1^x\sigma_3^x+h_x^2\sigma_2^x\sigma_3^x+h_z^2\sigma_1^z\sigma_2^z+(1+h_z^2)\sigma_1^z\sigma_3^z+h_z^2\sigma_2^z\sigma_3^z\nonumber\\
&\quad+h_xh_z(\sigma_1^x\sigma_2^z+\sigma_1^x\sigma_3^z+\sigma_2^x\sigma_1^z+\sigma_2^x\sigma_3^z+\sigma_3^x\sigma_1^z+\sigma_3^x\sigma_2^z)\nonumber\\
&\quad+h_x\sigma_1^x\sigma_2^z\sigma_3^z+h_x\sigma_1^z\sigma_2^z\sigma_3^x+2h_z\sigma_1^z+2h_z\sigma_2^z+2h_z\sigma_3^z+2h_z\sigma_1^z\sigma_2^z\sigma_3^z\big),\nonumber\\
\tr_{\bar A}H/d&=(\sigma_1^z\sigma_2^z+\sigma_2^z\sigma_3^z+h_x\sigma_1^x+h_x\sigma_2^x+h_x\sigma_3^x+h_z\sigma_1^z+h_z\sigma_2^z+h_z\sigma_3^z)/d_A,\nonumber\\
\tilde v&=\frac{\tr(H^2)}{Nd}=1+h_x^2+h_z^2,\quad\frac{\tr(H^3)}{Nd}=6h_z^2
\end{align}
so that
\begin{multline}
d_A\tr(X_A^2)=24h_x^2h_z^2+8h_x^2+12h_x^4+4(1+h_z^2)^2+16h_z^2\\
+\frac{36h_z^4\times3h_x^2}{(1+h_x^2+h_z^2)^2}+2h_z^4\left(\frac6{1+h_x^2+h_z^2}-2\right)^2+3h_z^2\left(\frac{6h_z^2}{1+h_x^2+h_z^2}-4\right)^2.
\end{multline}
Substituting $c$ from (\ref{eq:c}) into (\ref{eq:inter}), we obtain
\begin{equation}
S(\psi_A^\infty)=\ln d_A-\left(\frac{v-\tilde v}{2N}\right)^2\frac{d_A\tr(X_A^2)}{2\tilde v^4}+O(1/N^3).
\end{equation}

\begin{figure}
\centering
\begin{tikzpicture}
\begin{axis}
[xlabel=system size $N$,
ylabel=quadratic factor $\alpha$,
xmin=19,
xmax=31,
ymin=0.8,
ymax=0.91]
\addplot[only marks]
coordinates{(20,0.894)(22,0.887)(24,0.860)(26,0.863)(28,0.856)(30,0.850)};
\draw[dashed](19,0.818415511422098341)--(31,0.818415511422098341);
\end{axis}
\end{tikzpicture}
\caption{The quadratic factor $\alpha$ as a function of the system size $N$. The dots are the numerical results reproduced from the inset of Fig.~6(a) in Ref.~\cite{ML24}. The dashed line is our theoretical result (\ref{eq:num}) in the thermodynamic limit. Although one cannot conclude whether the dots approach the dashed line as $N\to\infty$, the trend looks promising.}
\label{fig}
\end{figure}
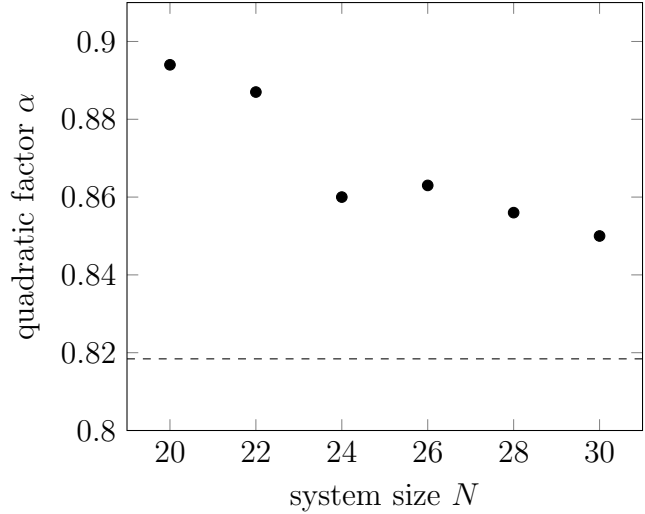

For $h_x=-1.05$ and $h_z=0.5$, $S(\psi_A^\infty)$ was computed for system sizes up to $N=30$ using exact diagonalization \cite{ML24}. Since Ref.~\cite{ML24} defines the von Neumann entropy using the binary logarithm, we convert its data by multiplying by $\ln 2$ for consistency with our convention of using the natural logarithm. For each $N$, the converted data are fitted by
\begin{equation}
S(\psi_A^\infty)=\ln d_A-\alpha\left(\frac{v-\tilde v}{2N}\right)^2.
\end{equation}
As shown in Fig.~\ref{fig}, the numerical values of the quadratic factor $\alpha$ semi-quantitatively support the theoretical prediction
\begin{equation} \label{eq:num}
\frac{d_A\tr(X_A^2)}{2\tilde v^4}\Big|_{h_x=-1.05,h_z=0.5}=\frac{568213558554560000}{694284933049739641}\approx0.818415511422098341.
\end{equation}

\section*{Acknowledgment}

I would like to thank Andreas M. L\"auchli and especially Ivo A. Maceira for helpful discussions on their numerical results. I also thank GPT-4o for its assistance in writing the abstract and introduction. This work was supported by Massachusetts unemployment insurance.

\bibliographystyle{unsrt}
\bibliography{main.bib}

\providecommand{\noopsort}[1]{}\providecommand{\singleletter}[1]{#1}%
\begin{thebibliography}{1}

\bibitem{Deu91}
J.~M. Deutsch.
\newblock Quantum statistical mechanics in a closed system.
\newblock {\em Physical Review A}, 43(4):2046--2049, 1991.

\bibitem{Sre94}
M.~Srednicki.
\newblock Chaos and quantum thermalization.
\newblock {\em Physical Review E}, 50(2):888--901, 1994.

\bibitem{RDO08}
M.~Rigol, V.~Dunjko, and M.~Olshanii.
\newblock Thermalization and its mechanism for generic isolated quantum
  systems.
\newblock {\em Nature}, 452(7189):854--858, 2008.

\bibitem{DKPR16}
L.~D'Alessio, Y.~Kafri, A.~Polkovnikov, and M.~Rigol.
\newblock From quantum chaos and eigenstate thermalization to statistical
  mechanics and thermodynamics.
\newblock {\em Advances in Physics}, 65(3):239--362, 2016.

\bibitem{Deu18}
J.~M. Deutsch.
\newblock Eigenstate thermalization hypothesis.
\newblock {\em Reports on Progress in Physics}, 81(8):082001, 2018.

\bibitem{ML24}
I.~A. Maceira and A.~M. L\"auchli.
\newblock Thermalization dynamics in closed quantum many body systems: a
  precision large scale exact diagonalization study.
\newblock arXiv:2409.18863.

\bibitem{main}
Y.~Huang.
\newblock High-precision simulation of finite-size thermalizing systems at long
  times.
\newblock arXiv:2406.05399.

\end{thebibliography}

\end{document}